\def\ha{{H$\alpha$}}
\def\hb{{H$\beta$}}

\def\LUM{\:{\rm ergs\:s^{-1}}}

\def\VEL{\:{\rm km\:s^{-1}}}

\def\OIGS{\:{\rm ergs\:cm^{-2}\:s^{-1}\:\AA^{-1}}}

\def\OiiiL{[\ion{O}{3}] $\lambda\lambda 4959, 5007$}
\def\SiiL{[\ion{S}{2}] $\lambda\lambda 6716, 6731$}
\def\NiiL{[\ion{N}{2}] $\lambda\lambda 6548, 6583$}
\def\OiL{[\ion{O}{1}] $\lambda\lambda 6300, 6363$}

\def\fe2{[\ion{Fe}{2}]}

\documentclass[12pt,preprint]{aastex}

\begin{document}


\newcommand{\MSOL}{\mbox{$\:M_{\sun}$}}  

\newcommand{\EXPN}[2]{\mbox{$#1\times 10^{#2}$}}
\newcommand{\EXPU}[3]{\mbox{\rm $#1 \times 10^{#2} \rm\:#3$}}  
\newcommand{\POW}[2]{\mbox{$\rm10^{#1}\rm\:#2$}}
\newcommand{\SING}[2]{#1$\thinspace \lambda $#2}
\newcommand{\MULT}[2]{#1$\thinspace \lambda \lambda $#2}
\newcommand{\CHINU}{\mbox{$\chi_{\nu}^2$}}
\newcommand{\vsini}{\mbox{$v\:\sin{(i)}$}}
\newcommand{\LSOL}{\mbox{$\:L_{\sun}$}}

\newcommand{\fuse}{{\it FUSE}}
\newcommand{\hst}{{\it HST}}
\newcommand{\iue}{{\it IUE}}
\newcommand{\euve}{{\it EUVE}}
\newcommand{\einstein}{{\it Einstein}}
\newcommand{\rosat}{{\it ROSAT}}
\newcommand{\chandra}{{\it Chandra}}
\newcommand{\xmm}{{\it XMM-Newton}}
\newcommand{\swift}{{\it Swift}}
\newcommand{\asca}{{\it ASCA}}
\newcommand{\galex}{{\it GALEX}}
\newcommand{\cxo}{CXO}


\shorttitle{An New Young Supernova Remnant in M83}
\shortauthors{Blair \etal}


\title{
{A Newly Recognized Very Young Supernova Remnant in M83}\altaffilmark{1,2,3}
}

\author{
William P. Blair\altaffilmark{4,14},
P. Frank Winkler\altaffilmark{5,11,14},
Knox S. Long\altaffilmark{6},
Bradley C. Whitmore\altaffilmark{6}, 
Hwihyun Kim\altaffilmark{7,8},
Roberto Soria\altaffilmark{9}, 
K. D. Kuntz\altaffilmark{4},
Paul P. Plucinsky\altaffilmark{10},
Michael A. Dopita\altaffilmark{11,12},
\&
Christopher Stockdale\altaffilmark{13}
}
\altaffiltext{1}{Based on observations obtained at the Gemini Observatory, which is operated by the Association of Universities for Research in Astronomy (AURA) under a cooperative agreement with the NSF on behalf of the Gemini partnership: the National Science Foundation (United States), the Science and Technology Facilities Council (United Kingdom), the National Research Council (Canada), CONICYT (Chile), the Australian Research Council (Australia), CNPq (Brazil) and CONICET (Argentina).}
\altaffiltext{2}{Based on observations with the NASA/ESA 
{\em Hubble Space Telescope}, obtained at the Space Telescope Science Institute, which is operated by the Association of Universities for Research in Astronomy, Inc., under NASA contract NAS5-26555.}
\altaffiltext{3}{Based on observations made with NASA's Chandra X-ray Observatory, which is operated by the Smithsonian Astrophysical Observatory under contract \# NAS83060, with data obtained through program GO1-12115.}
\altaffiltext{4}{The Henry A. Rowland Department of Physics and Astronomy, Johns Hopkins University, 3400 N. Charles Street, Baltimore, MD, 21218; wpb@pha.jhu.edu}
\altaffiltext{5}{Department of Physics, Middlebury College, Middlebury, VT, 05753; winkler@middlebury.edu}
\altaffiltext{6}{Space Telescope Science Institute, 3700 San Martin Drive, Baltimore, MD, 21218;  long@stsci.edu, whitmore@stsci.edu}
\altaffiltext{7}{Korea Astronomy and Space Science Institute, Daejeon 305-438, Republic of Korea; hwihyun@kasi.re.kr}
\altaffiltext{8}{Visiting Research Fellow, Department of Astronomy, The University of Texas at Austin, TX 78712, USA; hwihyun@astro.as.utexas.edu}
\altaffiltext{9}{International Centre for Radio Astronomy Research - Curtin University, GPO Box U1987, Perth, WA 6845, Australia; roberto.soria@icrar.org}
\altaffiltext{10}{Harvard-Smithsonian Center for Astrophysics, 60 Garden Street, Cambridge, MA 02138; plucinsky@cfa.harvard.edu }
\altaffiltext{11}{Research School of Astronomy and Astrophysics, Australian National University, Canberra, ACT 2611, Australia.}
\altaffiltext{12}{Astronomy Department, King Abdulaziz University, P.O. Box 80203, Jeddah, Saudi Arabia.}
\altaffiltext{13}{Department of Physics, Marquette University, PO Box 1881, Milwaukee, WI 53201-1881, USA; christopher.stockdale@marquette.edu}
\altaffiltext{14}{Visiting Astronomer, Gemini South Observatory, La Serena, Chile}

\begin{abstract}
As part of a spectroscopic survey of supernova remnant candidates in M83 using the Gemini-South telescope and GMOS, we have discovered one object whose spectrum shows very broad lines at  \ha, \OiL, and \OiiiL, similar to those from other objects classified as `late time supernovae.'  Although six historical supernovae have been observed in M83 since 1923, none were seen at the location of this object. 
{\it Hubble Space Telescope} Wide Field Camera 3 images show a nearly unresolved emission source, while \chandra\ and ATCA data reveal a bright X-ray source and nonthermal radio source at the position.  
Objects in other galaxies showing similar spectra are only decades post-supernova, which raises the possibility that the supernova that created this object occurred during the last century but was missed. 
Using photometry of nearby stars from the \hst\ data, we suggest the precursor was at least 17 \MSOL, and the presence of broad \ha\ in the spectrum makes a type II supernova likely.
The supernova must predate the 1983 VLA radio detection of the object.  We suggest examination of archival images of M83 to search for evidence of the supernova event that gave rise to this object, and thus provide a precise age.

\end{abstract}


Subject Headings: galaxies: individual (M83) -- galaxies: ISM  -- ISM: supernova remnants -- supernovae: general

Facilities:  Gemini-South Observatory (GMOS), Hubble Space Telescope (WFC3), Chandra X-ray Observatory

\section{Introduction \label{sec_intro}}

The transition from fading supernova (SN) to young supernova remnant (SNR) is not well characterized either observationally or theoretically \citep{fesen01}.  For a handful of objects where the SN was observed and there have been observations after the SN light has faded, so-called `late time' observations of the SN positions reveal very broad emission lines due to the rapidly expanding ejecta \citep[][and references therein]{{mili09}, {mili12}}.  The visibility of these broad lines at late times  seems to be restricted to core collapse SNe of various types. Even then, only a small subset of these SNe are detected at late times, depending on the mass loss history of the progenitor and/or the conditions in the surrounding interstellar or circumstellar medium.  

By way of comparison, spectra of young SNRs for which the SN was not observed can show essentially identical spectra, with very broad lines due to rapidly expanding ejecta for up to a few thousand years after the explosion.  Examples of these ejecta-dominated SNRs include the extraordinary young SNR in NGC~4449 \citep{mili08}, thought to be of order 70 years post-explosion \citep{bieten10}, Cassiopeia A in our Galaxy (see Fig.~14 of \citet{mili12} for an `integrated' spectrum of this object), which dates to $\sim$1680 AD, and even 1E0102-7219 in the Small Magellanic Cloud, which is still ejecta-dominated even $\sim$2000 years post-explosion \citep{{blair00}, {finkel06}, {vogt10}}.  

Because there are relatively few objects available to study this evolutionary phase and their characteristics are varied, we choose to combine these two groups together, and consider them all to be young SNRs.   This makes sense from the perspective that there may be SNe that are less than 100 years post-explosion but for which the SN event was not observed (like the NGC~4449 object mentioned above), which makes the distinction rather artificial.  However, even combining these groups, there are relatively few objects available to study.  
Thus, finding additional members of the class is one way of improving our understanding of the parameter space controlling this transition from fading SN to young SNR.

\begin{figure}
\plotone{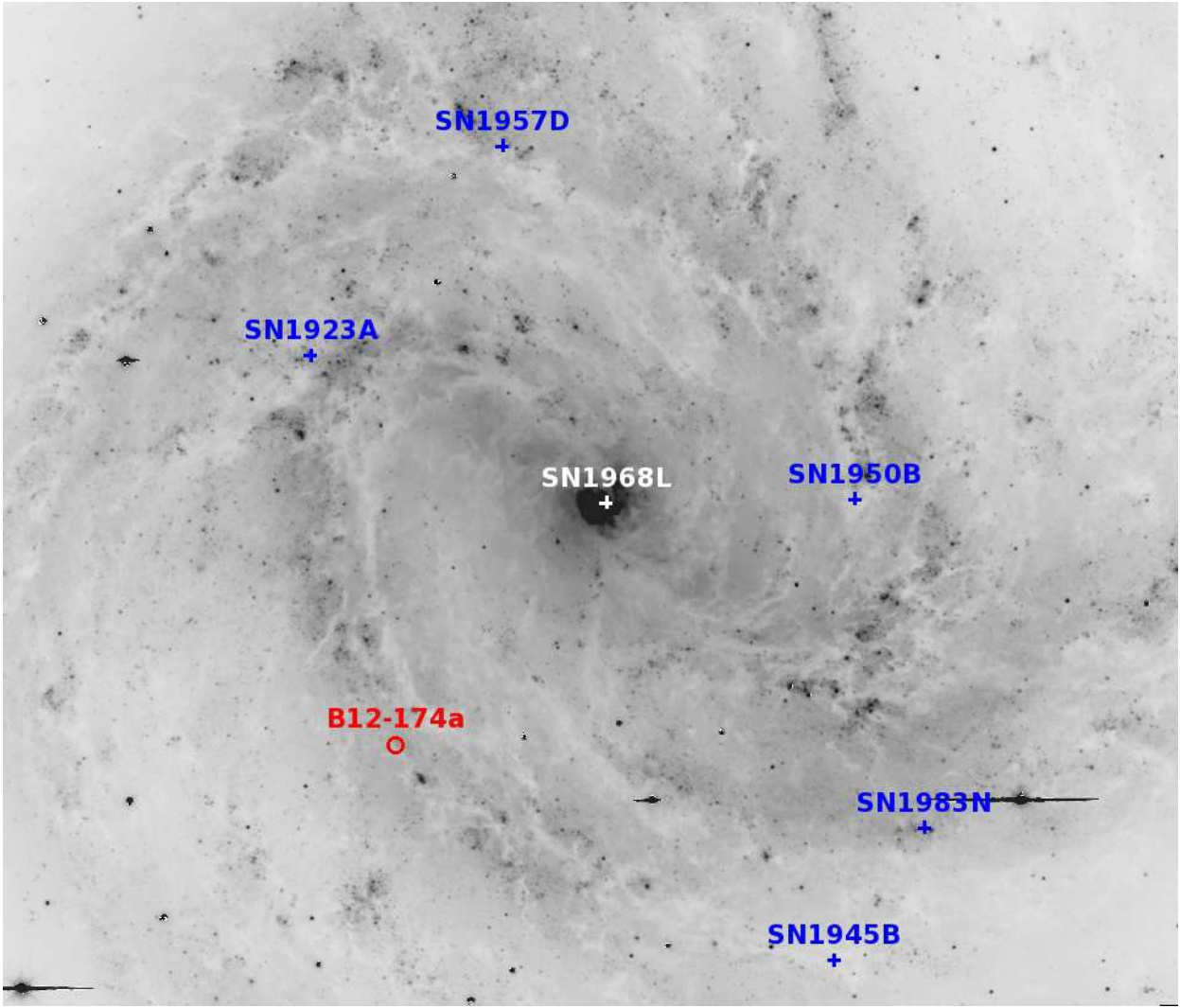}
\caption{A V-band image of M83 from the Magellan 6.5 m telescope with the positions of historical SNe marked as plus signs.  The location of B12-174a, the object of interest to this paper, is shown by a red circle in the arm to the SE of the starburst nucleus.  In this and all imaging figures that follow, north is up and east to the left.  \label{fig_sne}}
\end{figure}

We are pursuing a multi-wavelength observational campaign on the southern face-on spiral galaxy M83 (NGC\,5236; SAB(s)c; $i$ = 24$^{\circ}$; $d$ = 4.61 Mpc; \citet{saha06}) to study a broad range of topics including especially the young SNR population and the associated star formation activity \citep{{long12}, {blair12}, {long14}, {blair14}}. M83 has hosted six recorded supernovae (SNe) since 1923 \citep{{cowan85}, {stockdale06}}, second in number only to another face-on spiral, NGC\,6946, with nine.\footnote{Interestingly, three of the nine SNe in NGC\,6946 have occurred just since the turn of the millennium: SN2002hh, SN 2004et, and SN 2008S. See the SN catalogs by \citet{barbon99} and \citet{lennarz12} for a full accounting in both galaxies.}  The positions of these SNe across the face of M83 are shown for reference in Fig.~\ref{fig_sne} along with the position of the object that is the subject of this paper.

Optically, SNRs have most often been identified through higher ratios of \SiiL\ with respect to \ha\ than are typically found in photoionized gas, a signature of shock-heated emission \citep{mc72}.  In the case of M83, \citet[][henceforth BL04]{blair04} detected 71 SNR candidates using interference filter imagery in these two emission bands and confirmed 20 of these objects with spectroscopy.   More recently, we have expanded the SNR candidate list to at least 225 objects using imagery from the Magellan I 6.5 m telescope and IMACS instrument \citep[][henceforth B12]{blair12}\footnote{See also \citet{blair13} for a correction to the flux scale reported in B12.}.  Furthermore, we have used {\it Hubble Space Telescope} Wide Field Camera 3 (WFC3) images to investigate 63 of the smallest diameter candidates \citep{blair14}, about half of which were newly detected in the \hst\ survey and many of which are also strong soft X-ray sources \citep{long14}.  Between the \hst\ and Magellan surveys, there are now more than 250 SNR candidates in M83.

We are carrying out a new spectroscopic survey using Gemini-South and the GMOS instrument to follow-up many of these SNR candidates and elucidate their properties (Winkler et al., in preparation; see also \citet{long12} who observed the young remnant of SN1957D in M83).  As part of this survey, we have obtained the spectrum of an object that shows surprisingly broad lines, especially at \ha.  The object corresponds to BL04-50 in our initial survey and B12-174 in our more recent study, but the object of interest here appears as a bright knot or condensation on the northern limb of the larger object.  Hereafter, we will refer to this bright knot as B12-174a.  In this paper, we report an analysis of this object using a Gemini South spectrum along with the data from our \hst, \chandra, and ATCA radio observations, and conclude that it likely represents the remains of a SN that may have occurred within the last century but was not reported.


\section{New Spectrum and Other Supporting Data \label{sec_obs}}

\begin{figure}
\plotone{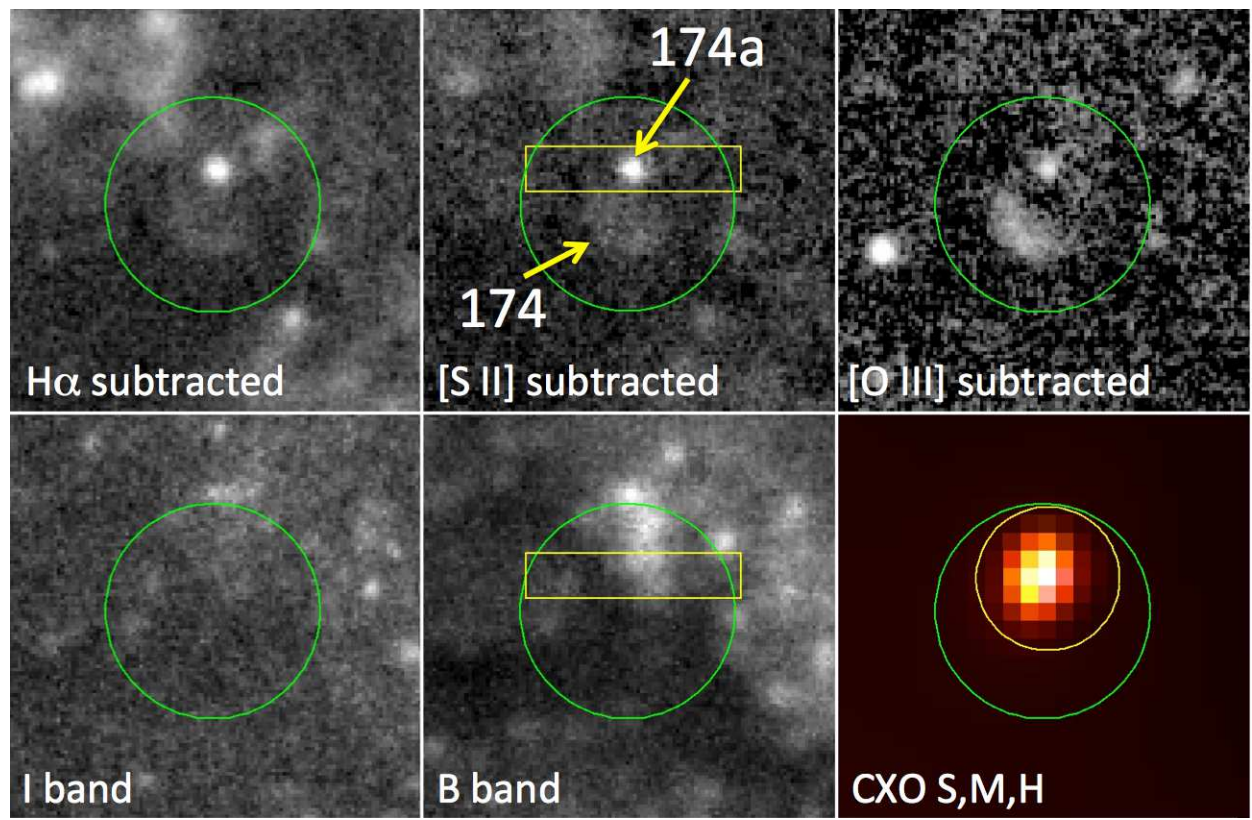}
\caption{This six-panel figure shows the region surrounding B12-174 in various bands. Panels a-e show Magellan IMACS imaging data as reported by \citet{blair12}, with panels a, b, and c showing continuum-subtracted H$\alpha$, [S~II], and [O~III] and panels d and e showing I-band and B-band.  Panel f shows a three-color representation of the \chandra\ X-ray data from \citet{long14}, with red being soft (0.35-1.1 keV), green being medium (1.1-2.4 keV), and blue being hard (2.4 - 8 keV) X-rays.  The green circle is 6\arcsec\ in diameter and is centered on the position given by \citet{blair12}, which assumed the southern arc of emission (labelled 174 in the [S~II] panel) was part of the same SNR.  However, the \chandra\ emission is clearly centered on the bright knot of emission that is offset toward the north, which we call B12-174a.  The yellow box indicates the position of the Gemini-GMOS 1\farcs25 $\times$ 6\arcsec\ aperture used to obtain the spectrum in Figure \ref{fig_spec}.   \label{fig_overview}}
\end{figure}

In Fig.~\ref{fig_overview} we show a six-panel context image of the region surrounding B12-174.  The optical data are from the Magellan imaging program described by B12, where the emission-line images have been continuum subtracted. I and B continuum bands are also shown, along with an aligned three-color image from the \chandra\ data reported by \citet[][object X316]{long14}.  
The green circle in the image is centered at the position given by B12, where the assessment of the object indicated that the bright knot toward the north was likely associated with a fainter extended shell of emission to the south and east that not only seemed to have elevated [S~II]/\ha\ emission but also emission in [O~III] $\lambda$5007.  Thus, the diameter of the overall object was measured to be 3\arcsec\ ($\sim$66 pc), and the object was not included in the \citet{blair14} assessment of the smallest diameter SNRs.  However, in Fig.~\ref{fig_overview} it is clear that the X-ray source is centered on the bright, unresolved optical knot that we now refer to as B12-174a, and it is this knot that we observed spectroscopically.  
Interestingly, the fainter extended emission identified as B12-174 seems to also show elevated [S II]/\ha\ along with detectable [O~III] emission, making it also likely to be shock excited.  It is tempting to suspect an association between these two nebulae, as is the case for the LBV nebula B416 in M33 \citep{fabrika05}, but a direct relation remains uncertain in the present case. 

\subsection{Spectroscopic Observations}

\begin{figure}
\plotone{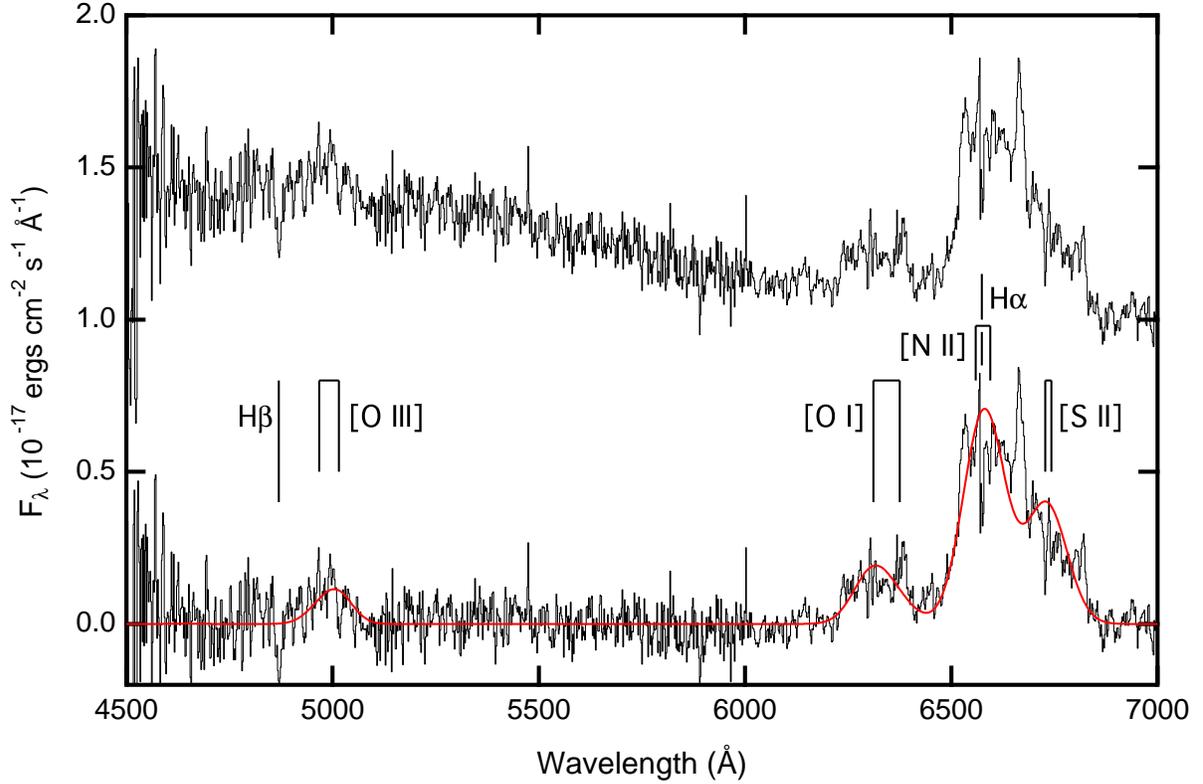}
\caption{The upper spectrum shows the calibrated Gemini-South GMOS spectrum of B12-174a, extracted from the bright point-like source in the yellow box shown in Fig.~\ref{fig_overview}, offset vertically by $0.7 \times 10^{-17} ~ \OIGS$ for clarity of presentation.  Very broad emission lines are seen superimposed on a blue continuum that is due to other stars within the aperture.   Below we show the same spectrum after fitting and subtracting the continuum.  The horizontal scale is observed wavelength, while line identifications are shown offset to M83 velocity (516 $\VEL$). The red curve superimposed on the lower trace represents a nominal fit using a single-component Gaussian for all ten indicated lines.  See text for details.  \label{fig_spec}}
\end{figure}

As part of a Gemini observing program (\# GS-2011A-C-1, Winkler, PI) to obtain spectra of SNRs in M83, we obtained a spectrum of B12-174a using the Gemini Multi-Object Spectrograph (GMOS) on the 8.2m Gemini South telescope on 2011 April 9 (UT).  Our observations consisted of  three 2000 s exposures, using the 600 line blue grating (G5323) and GG455 cutoff filter, in multi-slit mode with a custom mask.  For B12-174a, the  slit was 1\farcs25 wide and 6\arcsec\ long, oriented E-W as shown in Fig.~\ref{fig_overview}, and covered the wavelength range 4500 - 7000 \AA.    
Reduction proceeded using standard  Gemini techniques within the IRAF\footnote{IRAF is distributed by the National Optical Astronomy Observatory, which is operated by the Association of Universities for Research in Astronomy (AURA) under cooperative agreement with the National Science Foundation.} environment, and flux calibration was achieved using a number of spectrophotometric standards observed during the same 3-night classical observing run.  We subtracted the sky in the two-dimensional spectrum based on regions with relatively low emission at both ends of the slitlet shown in Fig.~\ref{fig_overview}, and then summed over the central 1\farcs2 to obtain the one-dimensional spectrum shown in the upper portion of Fig.~\ref{fig_spec}.

The spectrum shows a blue continuum with several very broad emission features superimposed.  As we will discuss below, the continuum is likely associated with blue stars within the aperture and is not directly related to the emission line source.  Hence, in the bottom portion of Fig.~\ref{fig_spec} we show a version of the spectrum where the continuum has been subtracted, leaving only the broad emission lines.  The dominant emission line covers the entire 6500 -- 6800 \AA\ range, while fainter broad lines are seen centered at roughly 6300 \AA\ and 5000 \AA.  From this spectrum, it is clear that the object was selected as having an elevated [S~II]/\ha\ ratio in the narrow band emission line imagery because both the narrow \ha\ and [S~II] filter bandpasses are contained within the broad $\sim$6650 \AA\ feature.  At 5000 \AA, the emission line is fainter, but because the entire narrow [O~III] imaging filter passband was filled with emission, the source nonetheless stands out well in continuum-subtracted [O~III] imaging data.

The $\sim$6650 \AA\ feature may be dominated by \ha, although it is broader than the other lines and shows an extension on the long-wavelength side that could be due to \SiiL.  Also, given the strength of \NiiL\ with respect to \ha\ observed for other M83 SNRs (see \citet{blair14}), the possible contribution of these lines cannot be ignored.  The structure at the rest position of \ha\ is a subtraction residual from an overlying narrow \ha\ component, but much of the structure in the broad line is likely to be real, indicating either a kinematically complex emitting structure or possibly an emitting clumpy shell (see \citet{danzig05}).

The broad $\sim$6300 \AA\ line is a blend of \OiL\
and should be dominated by the shorter wavelength line, which is normally three times stronger.  However, the line profile shows a nearly flat-topped structure with a narrower peak toward the long wavelength side, presumably indicating velocity substructure in the emitting material.
\footnote{For more information on observed [O~I] profiles in late time SNe and young SNR spectra and their variation with time, see \citet{mili10}.}  
The strongest (short wavelength) portion of the broad 6650 \AA\ feature shows a similar structure to [O~I], which may provide some indirect evidence that the long wavelength extended structure in this line is due to a separate feature, presumably broad \SiiL\ (see below).

The broad line at 5000 \AA\ is likely due to  \OiiiL, which again is normally seen in a 3:1 ratio but with the longer wavelength line dominating.  This broad [O~III] feature is faint enough that any expected substructure is lost in the noise, but we can say that the feature looks less flat-topped than the other lines ([O~I] in particular) and no narrow peak is seen on the long wavelength side, indicating possible differences in the kinematics and sub-structure of the [O~III]-emitting material compared with the other lines. No obvious broad \hb\ is visible at the current detection level.  

Because of the uncertainties caused by line blending and unknown relative line intensities, we have chosen a simplified approach to constraining the intrinsic line widths that may be present.   We have used the {\tt specfit} task in IRAF to fit a single-component Gaussian for each of the ten lines indicated in Fig.~\ref{fig_spec}.  The {\tt specfit} task varies allowed parameters systematically and uses an associated error array to minimize $\chi^{2}$ \citep{kriss94}.   We have fixed the central wavelengths  of all lines relative to \ha\ and fixed the relative intensities of the [N~II], [O~I], and [O~III] doublets to values dictated by atomic physics, but all other relative intensities were allowed to vary.    The red curve superimposed on the lower trace in Fig.~\ref{fig_spec} represents a nominal best fit using a single-component Gaussian for all ten lines, where all lines have a width of 5200 $\VEL$.   While the fit to the line profiles is not unique, this experiment provides a reasonable assessment of the broad line widths and lends credibility to the idea that the red ``wing" on the strong \ha + [N~II] blend is likely due to [S~II] with a similar velocity profile rather than an extended red wing of \ha\ emission.  This fit also highlights some of the most obvious non-Gaussian structure in the broad lines, including the relatively narrow spikes centered at $\sim$6383 \AA\ and 6662 \AA.  Because of the possibility of line blending, it is not clear whether these two features correlate directly in velocity space or not.

\subsection{Context from High-Resolution Imaging}

\begin{figure}
\plotone{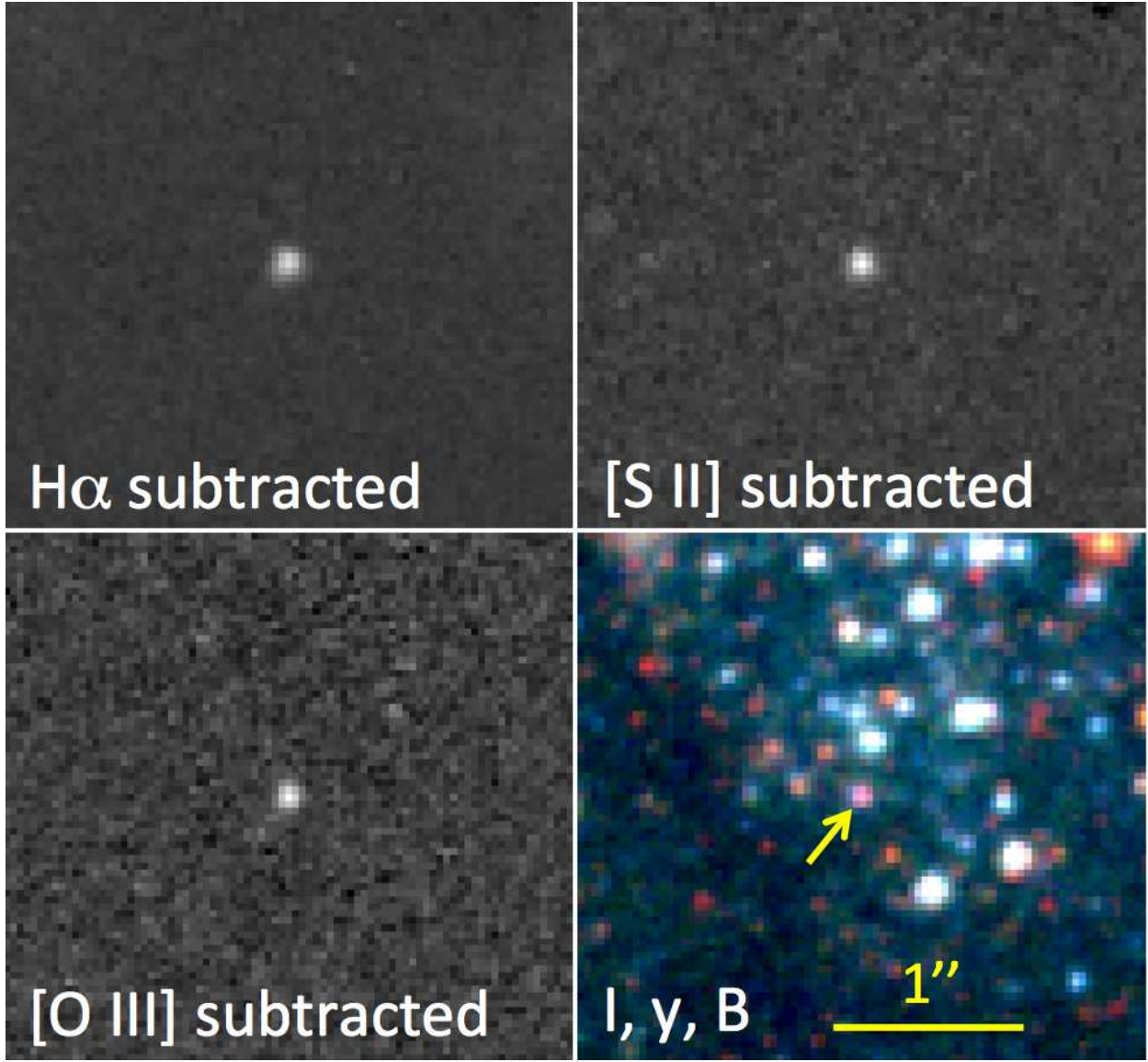}
\caption{The \hst\ WFC3 data for a 3\arcsec\ region centered on B12-174a.  The four panels show the continuum-subtracted \ha, [S~II], and [O~III] data, plus a three-color rendition of the \hst\ data, where RGB = I-band, narrow V (y) band, and B-band emission.  An unresolved point source is seen in the emission line images. The faint southern arc of B12-174 seen in Fig.~\ref{fig_overview} is too low in surface brightness to show up in the \hst\ data at this exposure level.  The yellow arrow indicates the faint red source at the same position as B12-174a.   \label{fig_wfc3}}
\end{figure}

We have inspected the \hst\ WFC3 imaging data for this region originally reported by \citet{blair14}.  The object is in our Field 3, and was imaged in a number of emission line and continuum filters, as reported in that paper.  Using the astrometric scale derived for these images, the position of B12-174a is RA(J2000) = 13:37:06.646; Dec(J2000) = $-$29:53:32.61, with an uncertainty we estimate as $\lesssim 0\farcs08$.
Fig.~\ref{fig_wfc3} shows a small, 3\arcsec\ region of the \hst\ imagery centered on B12-174a.  Indeed, the source is nearly unresolved in the subtracted emission line images, appearing essentially the same size as faint stars in the continuum panel of the Figure.  (However, see below for a more quantitative assessment.) The southern extended emission region B12-174 seen in the Magellan images is too low in surface brightness to be detected in the current WFC3 data.  

We have used the {\tt imexamine} task in IRAF to quantify the character of the emission line source {\it vis a vis} stars in the \hst\ data.  Averaging the results for the three emission line bands (taking cuts in both rows and columns of the images), we find a FWHM of 2.93 pixels for B12-174a, while averaging a handful of stars of comparable brightness in the F547M continuum exposure results in a FWHM of 2.49 pixels for the stars.  
Formally deconvolving this in quadrature is consistent with an intrinsic size of 1.18 pixels (FWHM) for B12-174a, which translates to 0.0466\arcsec\ or 1.02 pc at the assumed distance.  Hence, it is clear the size of the object must be quite small.

\begin{figure}
\plotone{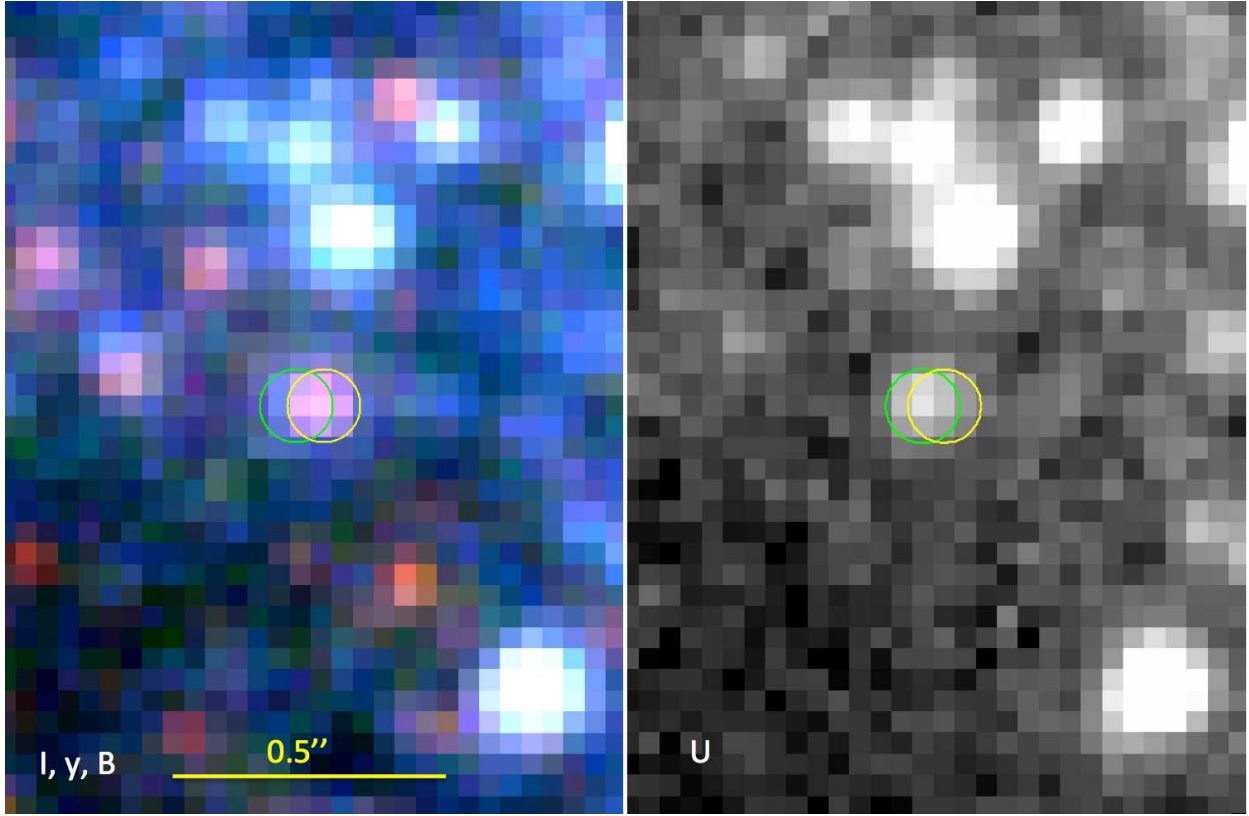}
\caption{An extreme close-up of the \hst\ continuum data near the B12-174a position.  At left is the same three-color image from Fig.~\ref{fig_wfc3}, and at right we show the F336W (U-band) image for comparison.  The green and yellow circles show the differing positions of the source in blue and red colors, implying a blue star lies in projection just to the east of the SNR. The centers of the two circles are separated by 1 pixel, or 0.04\arcsec. \label{fig_wfc3cont}}
\end{figure}

Interestingly, in the lower right panel of Fig.~\ref{fig_wfc3}, a faint red (I-band) source is coincident with the emission line source in the other panels. While this could be a red star, it seems probable, from comparison with spectra of similar objects in other galaxies \citep{mili12}, that some of this red emission arises from broad [O~II] $\lambda$7325 emission (and possibly fainter [Ca~II] and [Ar~III] emission seen in some other objects) within the F814W filter, which turns on sharply at 7000 \AA.  Without spectral coverage, however, this remains speculative. In Fig.~\ref{fig_wfc3cont} we provide an expanded view of the region showing the same color image from Fig.~\ref{fig_wfc3} on the left versus the U-band image on the right.  The yellow circle indicates the position of the emission line source, and coincides with the red source, while the green circle, offset one pixel (0.04\arcsec) to the east, is centered on the blue source in the U-band image.  Hence, a faint blue star lies very close along the line of sight.  
We note that the vertical extent of this Figure is 1\farcs25, the same as the GMOS slitlet; hence, it is highly probable that much or all of the blue continuum seen in the spectrum in Fig.~\ref{fig_spec} arises from the nearby blue stars visible in Fig.~\ref{fig_wfc3} and Fig.~\ref{fig_wfc3cont}, and not from B12-174a.

\subsection{X-ray and Radio Context}

B12-174a is coincident with the X-ray source X316 and the radio source A080 from \citet{long14}.   The positions listed by \citet{long14} for the \chandra\ and ATCA sources are both consistent with our \hst\ position to within 0.2\arcsec.  In that paper we list a 0.35 -- 8 keV count rate of 0.343 $\pm$ 0.024 $\times ~ 10^{-3}~ \rm cts~s^{-1}$ and a model-dependent flux of  F(0.35 - 8 keV) = 1.280 $\pm$ 0.083   $\times$ \POW{-6}{photons~cm^{-2}~s^{-1}}, with a corresponding luminosity of $L_x$ = 7.27~$\pm$~0.47 $\times$ \POW{36}{\LUM}.  The source has hardness ratios of (M $-$ S)/T = $-$0.50 $\pm$ 0.08 and (H $-$ M)/T = $-$0.06 $\pm$ 0.05, where the bands are defined as S =``soft'' = 0.35 -- 1.1 keV, M = ``medium'' = 1.1 -- 2.4 keV,  H = ``hard'' = 2.4 -- 8 keV, and T indicates Total.
This places the object well within the regime of other known SNRs although somewhat harder than average (see, for example, \citet{long14} Fig.~5 and 6).
The X-ray source was also listed in the earlier X-ray paper by \citet{soria03}, their source 100.  The somewhat higher count rate in these earlier \chandra\ data is offset by the changing response between 2000 and 2011 such that there is no evidence that the source has varied over the intervening 11 years.

A closer inspection of the \chandra\ data from \citet{long14}  shows a total of 316 counts (0.35--8.0 keV, 272.8 source counts and 
43.2 background counts) over the total 729 ks data set.  We used {\tt ACIS extract} \citep{broos10} to extract the X-ray spectrum and associated response files as described by \citet{long14}.  We fit the source and background spectra simultaneously in XSPEC version 12.8.1 using common detector and sky background models for both spectra.  We fit the unbinned spectra and used the C statistic to avoid the well known bias in $\chi^2$ when there are only a few counts per spectral bin.  
Once the best-fit parameters had been determined using the C statistic, we computed $\chi^2$ weighted by the model to provide an estimate of the goodness of fit.  Given the low number of counts, we fit the data with single component models for the source.  We adopted the \citet{wilms00} model for interstellar absorption  ({\tt tbabs} in  {\tt XSPEC}) and fit with both a nonthermal  ({\tt powerlaw}) model and a thermal model ({\tt APEC}; see \citet{foster12}) in {\tt XSPEC} \citep{arnaud96}.  

The best fit with the nonthermal model resulted in a C statistic of 744.6 with 1043 degrees of freedom, and a reduced $\chi^2$ of 1.05.  The best-fitted value of the ${\rm N_H}$ was $7.2\times 10^{20}~{\rm cm^{-2}}$ and the best fitted value of the photon power law index was 2.15. The Galactic column along the line of sight to M83 is $\sim4.0\times 10^{20}~{\rm cm^{-2}}$, so this derived ${\rm N_H}$ value seems consistent with a modest additional column intrinsic to M83. 

Left unconstrained, our thermal model fits wanted to drive the fitted ${\rm N_H}$ value to 0.0.  Hence, we forced a minimum value of ${\rm N_H}$ of $\sim4.0\times 10^{20}~{\rm cm^{-2}}$.   The best fit  thermal model resulted in a C statistic of 767.1 with 1043 degrees of freedom, and a reduced $\chi^2$ of 1.02.  In this model the ${\rm N_H}$ was indeed the minimum value that we allowed in the fit, and the best fitted value of the temperature was a surprisingly high kT=4.21~keV.  Given the already low value of the reduced $\chi^2$, we did not explore multi-component models as they are not justified by the quality of the data.  

\begin{figure}
\plotone{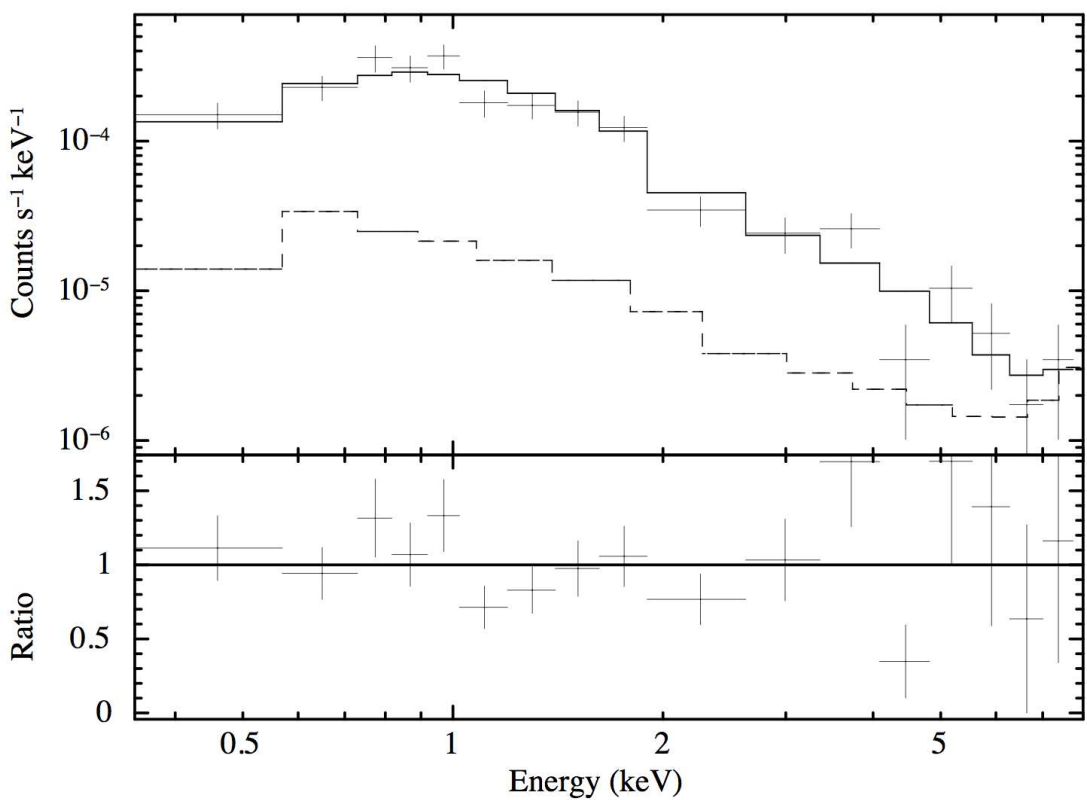}
\caption{An X-ray power law spectral fit for X316, the X-ray source that aligns with B12-174a.  The data are indicated by the crosses in the upper panel and the histogram represents the model fit to the source.  The source model includes a nonthermal component (a power law), a sky background component, and a detector background component.  The lower, dashed histogram is the model fit to the background spectrum that includes sky and detector background. The data have been binned for display purposes only. \label{fig_xmod}}
\end{figure}

Statistically, both the nonthermal and thermal models provide acceptable fits to the data.  For reference, the nonthermal model fit to the spectrum is shown in Figure~\ref{fig_xmod}.  The data and source plus background model are shown along with the background (sky+detector) model alone to demonstrate at which energies the source spectrum dominates the background spectrum. The data have been binned for display purposes only.   The nonthermal model provides a good fit to the data with no systematic pattern in the residuals.  We consider the nonthermal model to be more realistic, given both the quality of the fit and the reasonable value of the derived photon power law index.  The fact that the thermal model prefers an unphysical column and the best-fit value of the temperature is rather high for a young SNR argues somewhat against this model.  However, the quality of the spectrum does not allow us to exclude the thermal model or a combination of nonthermal and thermal components.  While the models are somewhat inconclusive, clearly the spectrum is relatively hard compared to those from most other young SNRs.
This might indicate the presence of a pulsar wind nebula or it might indicate an unusually high temperature for the object. We note a similar conclusion was reached for SN1957D by \citet{long12}, although an even harder spectrum and a much higher column density was inferred in that case. 

The source also corresponds with ATCA source A080 from \citet{long14}, and had been catalogued previously by \citet{maddox06} as source 41. \citet{maddox06} report 6 cm (4.873 GHz) and 20 cm (1.446 Ghz) VLA observations from 1998 and reprocess earlier VLA observation from 1983-84 \citep[][6 cm only]{cowan85} and 1990 (6 cm and 20 cm) for comparison.  The 20 cm flux in the latter two epochs remains nearly constant while the 6 cm flux drops rather dramatically over all three epochs, from 0.75 mJy (1983-84) to 0.42 mJy (1990) to 0.31 mJy (1998).  Comparing the \citet{maddox06} 4.873 GHz flux of 0.31 $\pm$  0.05 mJy in 1998 to the ATCA 5 GHz flux of 0.19 $\pm$ 0.02 mJy in 2011 appears to indicate that the earlier trend of declining flux has continued at $\sim$6 cm.  \citet{maddox06} also show the steepening spectral index from $-$0.28 in 1990 to $-$0.50 in 1998, although with substantial uncertainties.

\section{Discussion \label{sec_discussion}}

\subsection{Broad Lines and Line Identifications}

The broad emission lines in the B12-174a spectrum are reminiscent of a small class of `late time supernovae' (or very young SNRs from our preferred perspective) that have been observed spectroscopically in other nearby galaxies.  The best studied of these may be SN1979C in M100 \citep{mili09}, but \citet{mili12} present data on a number of other core-collapse SNe a few decades post-explosion, including SN1970G in M101, SN1980K in NGC\,6946, SN1993J in M81, and others.  Of the available comparisons, the one whose spectrum most resembles that of B12-174a is SN1970G in M101.  Although the 2010 spectrum of this object (Fig.~6 of \citet{mili12}) is noisy, it shows broad [O II]$ \lambda$7325, which may also be similar to B12-174a if some or all of the I-band emission noted previously is confirmed to be due to [O~II].  A number of the other objects shown by \citet{mili12} also have strong [O~II] $\lambda$7325. However, the class is small and varied, and observed characteristics likely depend on both age and environment. Thus, it is not clear that this similarity to SN 1970G can be used as an age indicator \citep[see][]{immler05}.

The shape of the broad feature near \ha\ (see Fig.~\ref{fig_spec}) is much broader and more asymmetrical than the two other lines of [O~I] and [O~III].    Our preferred explanation based on the spectral fits discussed above is to assume that the extension on the long wavelength side arises from [S~II] instead of \ha.  Making this assumption allows a single assumed width to apply to all of the lines, rather than assuming a highly asymmetrical \ha\ line. This red side extension is not always seen in the spectra of other similar objects, although a similar shape is present in SN1980K's spectrum (see Fig.~8 of \citet{mili12}).  

On the other hand, it is not even certain that the 6650 \AA\ feature is just \ha.  Super-solar abundances are seen in H~II regions across the bright optical disk of M83 \citep{bres02}, and many other M83 SNRs with spectroscopy show [N~II] $\lambda$6583 to be significantly stronger than \ha\ \citep{{blair04}, {blair14}}.  The same spectra show the sum of both [S~II] lines to often be comparable to or stronger than \ha.  Hence, if the red side extension is really [S~II], it could be an indirect indication that a significant portion of the broad feature centered near \ha\ is attributable to [N~II] emission in addition to \ha.  The spectral fits discussed earlier were consistent with the [N~II] lines contributing significantly to this feature, although the line blending does not allow hard constraints. 

The presence of [N~II] and [S~II] might also help explain the absence of obvious broad \hb\ in the spectrum shown in Fig.~\ref{fig_spec}:  If \ha\ dominates the 6650 \AA\ peak, significant extinction would be needed to reduce \hb\ below detectability.  This is marginally at odds with the low column densities indicated by the X-ray fits and the generally blue appearance of many of the stars in the region.  If a significant portion of the 6650 \AA\ feature is attributed to [N~II], a lower extinction would apply.  A higher quality spectrum including a more stringent constraint on any broad \hb\ feature would help elucidate this situation.

\subsection{A Very Young SNR?}

If we assume a diameter of 1 pc (consistent with the deconvolved source size inferred for the emission source) and an expansion velocity of 5200 $\VEL$ with no deceleration, we derive an age of about 96 years for the SNR.  Since many SNe start with much higher velocities than 5200 $\VEL$, it is possible that significant deceleration has taken place, in which case the object could be significantly younger than 100 years.  If the object has a diameter smaller than we estimate from our deconvolution, then again the object could be significantly younger than 100 years.  Given that many of the other members of the class are demonstrably much younger than 100 years, we consider it likely that B12-174a arose from a SN that occurred during the last century but which was not observed directly at the time of the explosion.  The detection of the object in the initial SNR survey of BL04, using data obtained in 1991 April, and the radio flux measurement at 6 cm by \citet{cowan85} from 1983 indicates that the SN must have occurred prior to 1983.  The rapid change in 6 cm flux between 1983 and 1990 and a slower decline after that may indicate a rapid change in local environment conditions near the SN position, possibly indicating a SN that took place not long before 1983.

From available SN catalog information \citep{{barbon99}, {lennarz12}}, the brightest observed SNe in M83, presumably caught near peak magnitude, are SN1968L (Type II, V = 11.9) and SN1983N (Type Ib, V  = 11.7).  A wide range of core collapse SN light curves show drops from peak by 3 -- 4 magnitudes over the first 100 days \citep{fili97}.  Assuming a SN occurred near the time the galaxy slid into the solar avoidance zone and was unobserved for 3 -- 4 months, such a SN could have dropped to V $\simeq$ 16 or fainter and easily have been missed, depending on the particular SN type and other circumstances.

Furthermore, a cursory assessment of the historical record readily shows significant incompleteness that gets more significant the earlier one goes.  Using the online version of the Asiago catalog \citep{barbon99}\footnote{See http://cdsarc.u-strasbg.fr/viz-bin/Cat?B/sn.}, making a cut at galaxies with velocities $\le$ 700 $\VEL$ (or roughly within 10 Mpc), excluding known type Ia SNe, and binning into 20 year bins, one sees the following number of SNe per indicated period:  1901-1920: 3; 1921-1940: 9; 1941-1960: 9; 1961-1980: 14; 1981-2000: 17; and 2001 - 2014: 27.  Hence, it seems reasonable that a SN in M83, especially during the period prior to $\sim$1980, could have been missed.

It would be important to pin down an exact time since the explosion for B12-174a if possible, and so we encourage a search of archival image data at the B12-174a position (see Fig.~\ref{fig_sne}) for any indication of the SN that created this young remnant.  
Unfortunately, if nothing is found, we will not be able to separate the idea that the SN predates the historical record from the possibilities that the SN was significantly sub-luminous compared to normal core collapse SNe or that it was simply missed.

\subsection{Clues to the Progenitor}

SN1957D, also in M83, forms an interesting comparison to B12-174a.  In the late-1980's a spectrum of this young remnant showed broad oxygen lines, in particular at [O~III] $\lambda$5007 \citep{long89}.  As reported by \citet{long92} and more recently by \citet{long12}, however, this broad emission has faded rather dramatically on a fairly short timescale. The hard X-ray source and radio characteristics of SN1957D make a plerion interpretation likely for this young remnant.  A major difference between SN1957D and B12-174a is the presence of broad \ha\ (and possibly [N~II]) in the spectrum of the latter, which hints strongly for a type II SN designation for the B12-174a SN.  SN1957D was discovered well past maximum and the SN type was not determined at the time.  

As described in detail by \citet{kim12}, careful photometry in multiple continuum bands can be used to investigate the extinction properties and characteristic age or range of ages in a particular stellar population.  \citet{long12} used \hst\ photometry and this CMD fitting technique to determine the characteristics of the stellar population near SN1957D.  Many of the stars were consistent with a very young isochrone of 4 Myr, but the envelope of points seemed to be bounded by the 10 Myr isochrone, indicative of a main sequence turn-off mass of at least 17 \MSOL\ (and possibly significantly higher) for the associated cluster of stars.  Hence, there is little doubt SN1957D was a core collapse SN of some type. 

\begin{figure}
\plotone{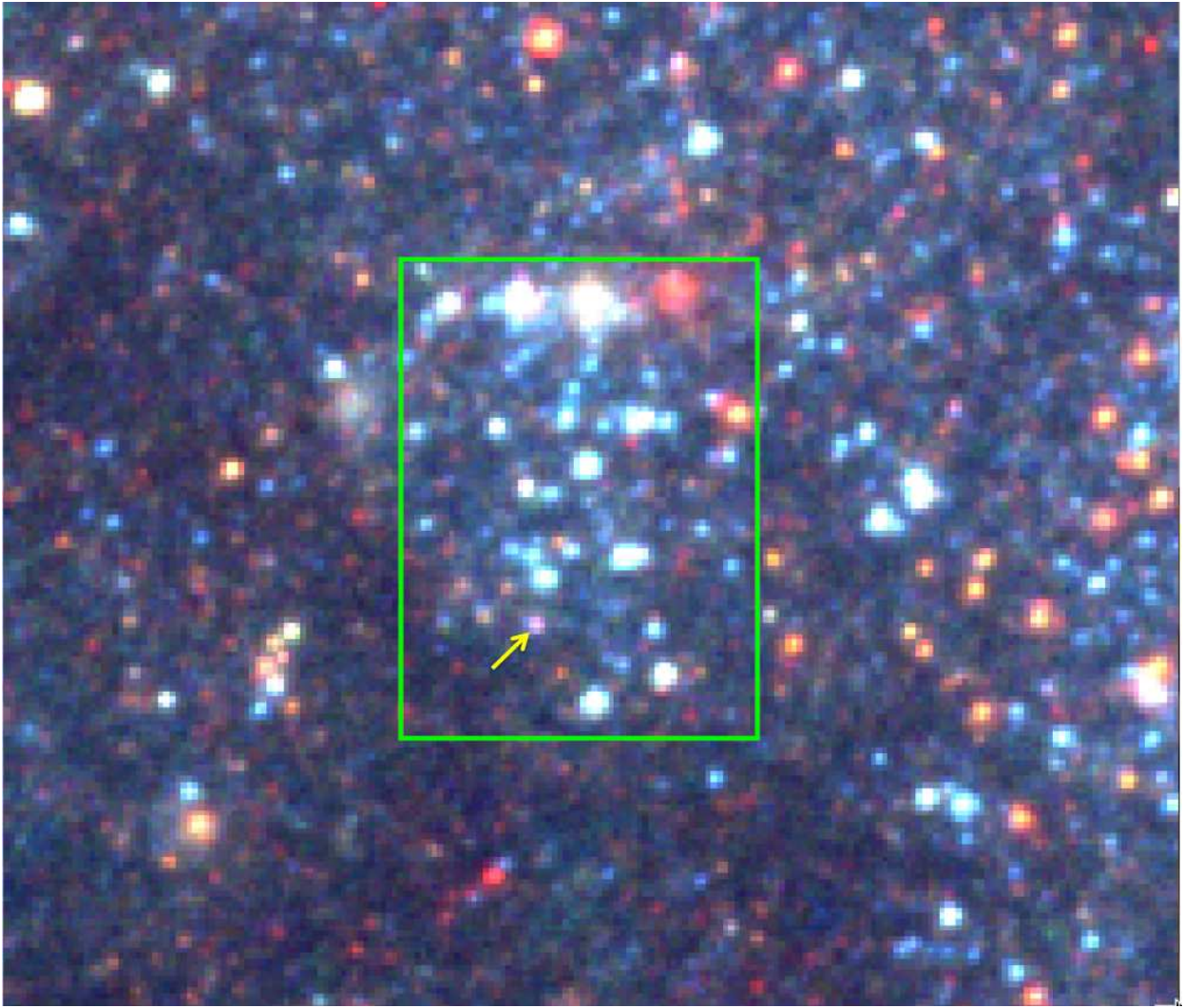}
\caption{A wider (8\farcs3 $\times$ 7\arcsec) view of the three-color \hst\ continuum image, indicating the 2\farcs5 by 3\farcs5 (or roughly 55 by 76 pc) region over which CMD fitting was performed (green box).  Excess extinction can be seen around the outside edges of the region, especially in the east and south,  The color scale is the same as used in previous Figures. The arrow shows the position of B12-174a. \label{fig_cmdreg}}
\end{figure}

We have carried out a similar photometric analysis to estimate the age and main sequence turnoff mass for the population near B12-174a, again using the procedures outlined by \citet{kim12}.  The object is located within a spiral arm to the SE of the M83 nucleus (see Fig.~\ref{fig_sne}), and appears to be associated with a well-defined grouping of stars, shown by the green box in Fig.~\ref{fig_cmdreg}.  The arrow in this Figure indicates the position of B12-174a for reference. We do not center the region of interest on the B12-174a position, but rather set the region by eye to be the grouping of bright stars that seem to be associated.  Taking a larger region would only include more `background' stars, while a significantly tighter region would just diminish the number of stars available for photometry.  The size of the selected region is $\sim$55 $\times$ 76 pc, which is reasonable for a young, coeval grouping.
 
\begin{figure}
\plotone{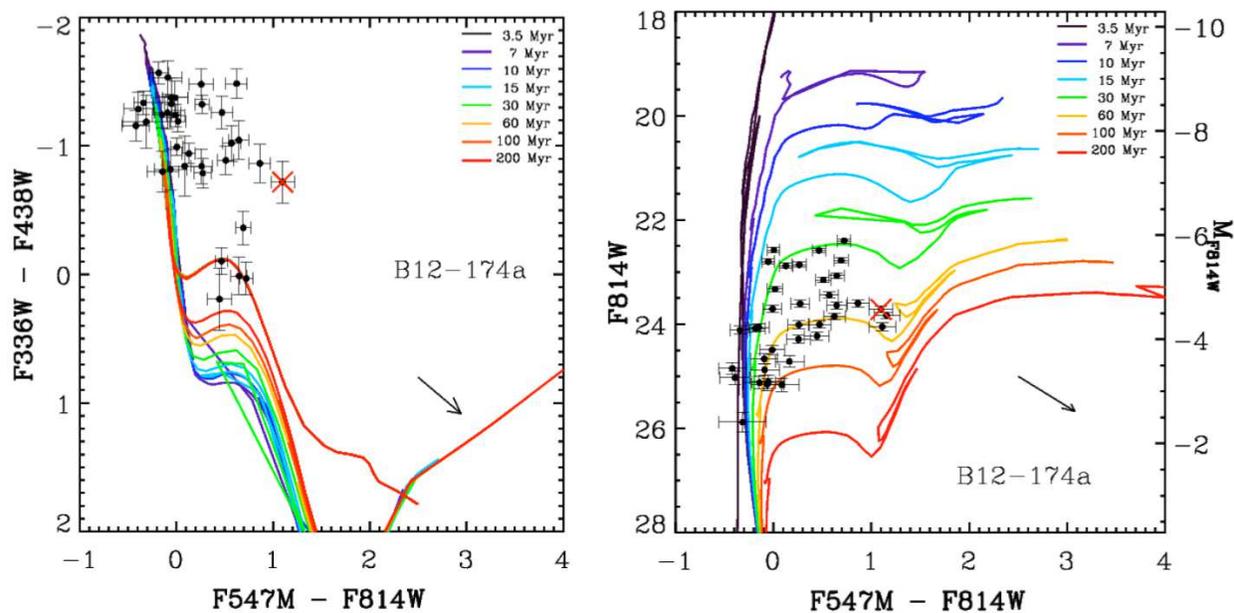}
\caption{Results of \hst\ aperture photometry for stars in the region surrounding B12-174a.  (left) A color-color diagram; (right) a color-magnitude diagram.  The curves are isochrones from the Padova models as described in the text.  Each dot is a star from within the rectangle shown in Fig.~\ref{fig_cmdreg} as observed (no correction for extinction). The arrows indicate the direction of motion of points due increasing extinction, assuming a standard R = 3.1 extinction curve.  The point marked with a red `X' is the source at the position of B12-174a. \label{fig_cmd1}}
\end{figure}

Aperture photometry was carried out in the \hst\ F336W (U), F438W (B), F547M (narrow-V), and F814W (I) broadband images\footnote{Although we do not actually transform to the Johnson--Kron--Cousins UBVI system, we will refer to these filters as U, B, V, and I for convenience of presentation in the text.} for 72 stellar objects in the region outlined in Fig.~\ref{fig_cmdreg}.  The results are shown in the color-color diagram (left) and CMD (right)  in Fig.~\ref{fig_cmd1}.   We only plot stars with photometric errors less than 0.25 mag in (V -- I) color and (V -- I) $<$ 1.2, as redder stars tend to be heavily extincted and do not have U and/or B colors that are required for the correction procedure (see section 2.4 of \citet{kim12}).  
 The lines shown in both panels of Fig.~\ref{fig_cmd1} are Padova stellar isochrones (CMD V2.3; http://stev.oapd.inaf.it/cmd; \citet{marigo08}) for the \hst\ WFC3 filters, Z = 1.5 $\rm Z_{\sun}$, and various ages (6.0 $\le$ log (age) $\le$ 9.95).  We applied a Galactic foreground extinction of $\rm A_V$ = 0.19 for M83 (from NED, \citet{schlaf11}).  The spread of the points in the color-color panel of Fig.~\ref{fig_cmd1} is largely due to differential internal extinction between individual stars. The arrow in each panel shows the direction of motion of points in each diagram with increasing reddening.

\begin{figure}
\plotone{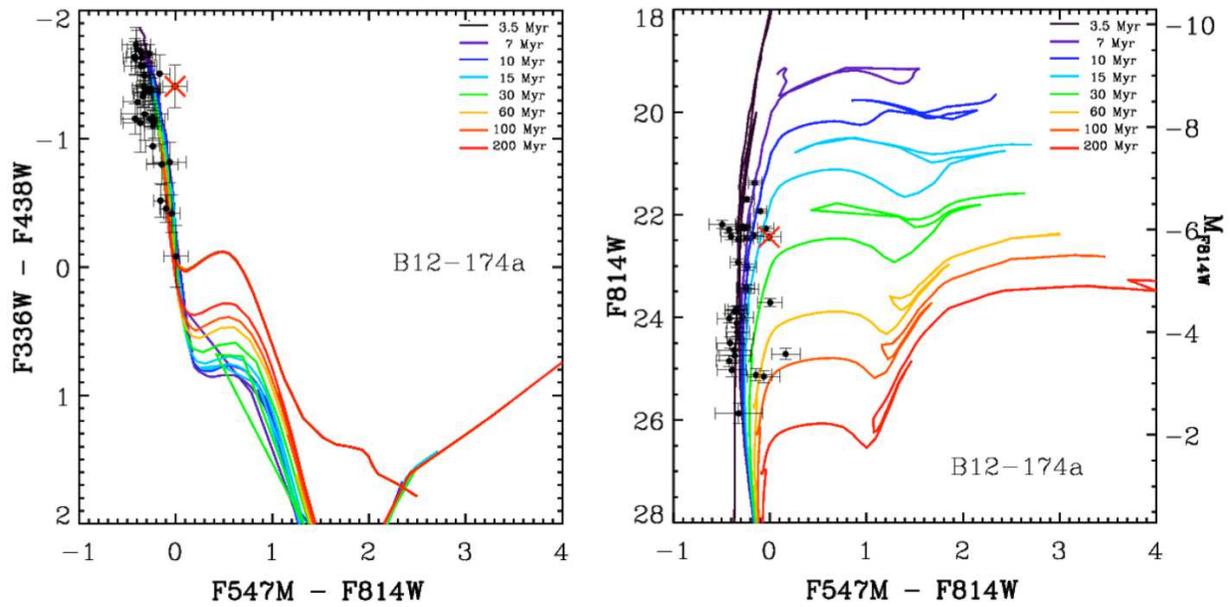}
\caption{Same as Fig.~\ref{fig_cmd1}, but after applying correction for extinction to each star, as described in the text.   Most of the stars lie on the isochrones for young stars, with modest scatter.  Again, the point marked with a red `X' is the corrected position of the source that is coincident with B12-174a.  It appears somewhat too red, consistent with I-band contamination by SNR emission, as described in the text.  \label{fig_cmd2}}
\end{figure}

Fig.~\ref{fig_cmd2} shows the same two plots after applying the star-by-star extinction correction. The technique is to unredden the stars in the color-color panel by moving them backwards along the reddening vector (toward upper left) until they match the isochrone-predicted values of the reddening-free Q parameter (see eqn.~1 in \citet{kim12}; also \citet{mihalas81}). The corrected colors and magnitudes are then used in the CMD to determine the intrinsic characteristics of the stars.   Again, details are provided in sec. 2.4 of \citet{kim12}.

There is some spread in the points around the youngest isochrones in Fig.~\ref{fig_cmd2}, but the data are best fit with the 7 -- 10 Myr curves, consistent with the group being old enough that no \ha\ is seen associated with the stars but still quite young.   This is very similar to the age derived for the association adjacent to SN1957D. Inspection of the isochrone data files indicates a nominal MS turnoff mass of 22 -- 17 \MSOL\ for 7 Myr and 10 Myr, respectively.  The brightest stars in the corrected CMD are clearly younger than the 15 Myr isochrone, so we consider $\sim$17 \MSOL\ to represent a reasonable lower limit to the B12-174a progenitor mass, again consistent with a core-collapse SN type.  The corrected position of the B12-174a source in this Figure is somewhat too red, which is consistent with the earlier suggestion that the I-band flux for this source may be contaminated by [O~II] $\lambda$7325 emission.

\section{Summary \label{sec_summary}}

Adding Gemini-South spectroscopy to earlier \chandra, \hst, radio, and ground-based optical data sets, we have discovered that the object B12-174a represents a very young remnant of a SN in M83 that likely occurred in the past century but was not reported.  The spectrum shows very broad lines at the positions of \ha, [O~I], and [O~III] and looks similar in many ways to spectra of a handful of `old supernovae'  in other nearby galaxies that date to only decades after the explosion (the oldest being SN1970G in M100).  A line-fitting exercise indicates an approximate expansion velocity near 5200 $\VEL$ as well as the presence of broad [S~II] and possibly [N~II] in addition to \ha. The presence of a radio source in data from 1983 implies a date prior to this for the SN explosion, and we encourage a search of archival images of M83 for evidence of any activity at the position of B12-174a.  Photometry of stars in the direct vicinity in the \hst\ data indicates a young population and a potentially massive precursor, and makes an association with a core collapse SN of type II likely.

Historically determined observational SN rates in nearby galaxies are of interest for constraining other more indirect determinations of SN rates, but are fraught with the problems of incompleteness and small number statistics.   The identification of B12-174a raises the possibility that careful surveys of nearby galaxies may be able to correct these rates for missed SNe and arrive at improved SN rates for galaxies such as M83. 

\acknowledgements

We thank additional colleagues on the companion multi-wavelength surveys of M83 (\chandra, ATCA, and \hst) for useful discussions.
WPB and PFW acknowledge travel support from the National Optical Astronomy Observatory, as well as STScI grants under the umbrella program ID GO-12513 to Johns Hopkins University and Middlebury College, respectively. Support was also provided by the National Aeronautics and Space Administration through \chandra\ grant number G01-12115, issued by the \chandra\ X-ray Observatory Center, which is operated by the Smithsonian Astrophysical Observatory for and on behalf of NASA under contract NAS8-03060. WPB acknowledges support from the Dean of the Krieger School of Arts and Sciences and the Center for Astrophysical Sciences at Johns Hopkins University during this work.  PFW acknowledges additional support from the National Science Foundation through grant AST-0908566.

\newpage

\bibliographystyle{apj}

\begin{thebibliography}{}

\bibitem[Arnaud(1996)]{arnaud96} Arnaud, K.~A.\ 1996, Astronomical Data Analysis Software and Systems V, ed. by G. Jacoby \& J. Barnes, 101, 17

\bibitem[{{Barbon} {et~al.}(1999){Barbon}, {Buond{\'{\i}}}, {Cappellaro}, \&
  {Turatto}}]{barbon99}
{Barbon}, R., {Buond{\'{\i}}}, V., {Cappellaro}, E., \& {Turatto}, M. 1999,
  \aaps, 139, 531; see also http://web.oapd.inaf.it/supern/cat/cat.txt
  

\bibitem[Bietenholz et al.(2010)]{bieten10} Bietenholz, M.~F., Bartel, N., Milisavljevic, D., et al.\ 2010, \mnras, 409, 1594

\bibitem[{{Blair} \& {Long} (2004) {Blair},{Long}}]{blair04} {Blair}, W.~P., \& {Long}, K.~S.\ 2004, \apjs, 155, 101 [BL04] 

\bibitem[Blair et al.(2014)]{blair14} Blair, W.~P., Chandar, R., Dopita, M.~A., et al.\ 2014, \apj, 788, 55

\bibitem[Blair et al.(2000)]{blair00} Blair, W.~P., Morse, J.~A., Raymond, J.~C., et al.\ 2000, \apj, 537, 667

\bibitem[Blair et al.(2012)]{blair12} Blair, W.~P., Winkler, P. F., \& Long, K. S. \ 2012, \apjs, 203, 8 [B12]

\bibitem[Blair et al.(2013)]{blair13} Blair, W.~P., Winkler, P. F., \& Long, K. S. \ 2013, \apjs, 207, 40

\bibitem[{{Bresolin} \& {Kennicutt}(2002)}]{bres02}{Bresolin}, F., \& {Kennicutt}, R.~C. 2002, \apj, 572, 838


\bibitem[Broos et al.(2010)]{broos10} Broos, P.~S., Townsley, L.~K., Feigelson, E.~D., et al.\ 2010, \apj, 714, 1582


\bibitem[Cowan \& Branch(1985)]{cowan85} Cowan, J.~J., \& Branch, D.\ 1985, \apj, 293, 400 


\bibitem[Danziger(2005)]{danzig05} Danziger, I.~J.\ 2005, Mem.S.A.It., 7, 82 


\bibitem[Fabrika et al.(2005)]{fabrika05} Fabrika, S., Sholukhova, O., Becker, T., et al.\ 2005, \aap, 437, 217 



\bibitem[Fesen (2001)]{fesen01} Fesen, R.~A. 2001, in Young Supernova Remnants, AIP Conf. Proc., 565, 119

\bibitem[Filippenko(1997)]{fili97} Filippenko, A.~V.\ 1997, \araa, 35, 309 

\bibitem[Finkelstein et al.(2006)]{finkel06} Finkelstein, S.~L., Morse, J.~A., Green, J.~C., et al.\ 2006, \apj, 641, 919

\bibitem[Foster et al.(2012)]{foster12} Foster, A.~R., Ji, L., Smith, R.~K., \& Brickhouse, N.~S.\ 2012, \apj, 756, 128 

     

\bibitem[Immler \& Kuntz(2005)]{immler05} Immler, S., \& Kuntz, K.~D.\ 2005, \apjl, 632, L99


\bibitem[Kim et al.(2012)]{kim12} Kim, H., Whitmore, B.~C., Chandar, R., et al.\ 2012, \apj, 753, 26

   
\bibitem[Kriss(1994)]{kriss94} Kriss, G. 1994, Astronomical Data Analysis Software and Systems III, 61, 437

\bibitem[Lennarz et al.(2012)]{lennarz12} Lennarz, D., Altmann, D., \& Wiebusch, C.\ 2012, \aap, 538, A120

\bibitem[Long et al.(1989)]{long89} Long, K.~S., Blair, W.~P., \& Krzeminski, W.\ 1989, \apjl, 340, L25 

\bibitem[Long et al.(1992)]{long92} Long, K.~S., Winkler, P.~F., \& Blair, W.~P.\ 1992, \apj, 395, 632 

\bibitem[Long et al.(2012)]{long12} Long, K.~S., et al.\ 2012, \apj, 756, 18

\bibitem[Long et al.(2014)]{long14} Long, K.~S., et al.\ 2014, \apjs. 212, 21

\bibitem[Maddox et al.(2006)]{maddox06} Maddox, L.~A., Cowan, J.~J., Kilgard, R.~E., et al.\ 2006, \aj, 132, 310

\bibitem[Marigo et  al.(2008)]{marigo08} Marigo, P., Girardi, L., Bressan, A., et al.\ 2008, \aap, 482, 883

\bibitem[Mathewson \& Clarke(1972)]{mc72} Mathewson, D.~S., \& Clarke, J.~N.\ 1972, \apjl, 178, L105

\bibitem[Mihalas \& Binney(1981)]{mihalas81} Mihalas, D., \& Binney, J.\ 1981, San Francisco, CA, W.~H.~Freeman and Co., 1981

\bibitem[Milisavljevic \& Fesen(2008)]{mili08} Milisavljevic, D., \& Fesen, R.~A.\ 2008, \apj, 677, 306

\bibitem[Milisavljevic et al.(2009)]{mili09} Milisavljevic, D., Fesen, R.~A., Kirshner, R.~P., \& Challis, P.\ 2009, \apj, 692, 839

\bibitem[Milisavljevic et al.(2010)]{mili10} Milisavljevic, D., Fesen, R.~A., Gerardy, C.~L., Kirshner, R.~P., \& Challis, P.\ 2010, \apj, 709, 1343

\bibitem[Milisavljevic et al.(2012)]{mili12} Milisavljevic, D., Fesen, R.~A., Chevalier, R.~A., et al.\ 2012, \apj, 751, 25

\bibitem[Saha et al.(2006)]{saha06} Saha, A., Thim, F., Tammann, G.~A., Reindl, B., 
\& Sandage, A.\ 2006, \apjs, 165, 108 

\bibitem[Schlafly \& Finkbeiner(2011)]{schlaf11} Schlafly, E.~F., \& Finkbeiner, D.~P.\ 2011, \apj, 737, 103


\bibitem[Soria \& Wu(2003)]{soria03} Soria, R., \& Wu, K.\ 2003, \aap, 410, 53 


\bibitem[Stockdale et al.(2006)]{stockdale06} Stockdale, C.~J., Maddox, L.~A., Cowan, J.~J., Prestwich, A., Kilgard, R., 
\& Immler, S.\ 2006, \aj, 131, 889 
  
\bibitem[Vogt \& Dopita(2010)]{vogt10} Vogt, F., \& Dopita, M. A.\ 2010, \apj, 721, 597 



\bibitem[Wilms et al.(2000)]{wilms00} Wilms, J., Allen, A., \& McCray, R.\ 2000, \apj, 542, 914



\end{thebibliography}

\expandafter\ifx\csname natexlab\endcsname\relax\def\natexlab#1{#1}\fi

\clearpage

\end{document}